\documentstyle[12pt]{article}
\begin{document}
\font\ninerm = cmr9

\def\footnoterule{\kern-3pt \hrule width \hsize \kern2.5pt}

\pagestyle{empty}

\begin{flushright}
gr-qc/0106005 \\
$~$ \\
January 2001
\end{flushright}

\vskip 0.5 cm

\begin{center}
{\large\bf Observed threshold anomalies as the first hope of a
manifestation of Planck-length physics\footnote{Talk given at 
the ``9th Marcel Grossmann Meeting",
2-8 July 2000, University of Rome {\it La Sapienza}
(to appear in the proceedings).}}
\end{center}
\vskip 1.5 cm
\begin{center}
{\bf Giovanni AMELINO-CAMELIA}\\
\end{center}
\begin{center}
{\it Dipart.~Fisica,
Univ.~Roma ``La Sapienza'',
P.le Moro 2, 00185 Roma, Italy}
\end{center}

\vspace{1cm}
\begin{center}
{\bf ABSTRACT}
\end{center}

{\leftskip=0.6in \rightskip=0.6in
The observations of photons from the BL Lac object Mk501
with energies above 10 TeV and of cosmic rays with 
energies above the GZK threshold appear to be inconsistent
with conventional theories.
Remarkably, among the recent new-physics proposals 
of solutions of these threshold paradoxes
a prominent role has been played by proposals based
on quantum properties of space-time.
While the experimental evidence (and theory work attempting to
interpret it) is much too preliminary to 
justify any serious hopes that we might have stumbled upon the first
manifestation of a ``quantum gravity", the fact that for
the first time phenomenological models involving quantum-gravity
ideas are competing on level ground with other new-physics proposals
clearly marks the beginning of a new stage of
quantum-gravity research.
I emphasize one important aspect of this new
phenomenology: combining the determination
of the relevant thresholds with data on the time/energy structure 
of gamma-ray bursts it is possible to distinguish between
alternative quantum-gravity scenarios.
This point is illustrated focusing on 3 specific scenarios:
dispersion-inducing space-time foam, string-theory-motivated
non-commutative space-time, and this author's recent proposal
of a relativistic theory in which the Planck length has the role
of fundamental observer-independent minimum length.}

\newpage
\baselineskip 12pt plus .5pt minus .5pt
\pagenumbering{arabic}
\pagestyle{plain} 

When Lee Smolin suggested that I contribute to the MG9 session
he was to chair a lecture giving an ``update" on the status
of Planck-length phenomenology\cite{polonpap,grf00},
I was somewhat concerned. Lee clearly was hoping 
I could discuss significant new developments
with respect to the status 
of Planck-length phenomenology described in my 1999
notes\cite{polonpap}, 
but this phenomenology 
(which was thought to be impossible until very recently) 
remains very challenging and its progress
should naturally have a very slow pace.
My fear that there would not be any significant new developments
for me to discuss proved to be unjustified when, just two
months before MG9, Protheroe and Meyer\cite{aus} 
pointed out that some puzzling observations of photons 
from the BL Lac object Mk501
with energies above 10 TeV 
could be explained using the 
Planck-length-phenomenology model
of Ref.~[4].
This analysis by Protheroe and Meyer combines with a previous analysis
by Kifune\cite{kifu}, which concerns another possibly related
paradox in astrophysics data,
in such a way that for the first time it appears legitimate to 
wonder whether some available (real-world!) data 
actually are a manifestation of Planck-length physics.
Also significant for the outlook of Planck-length phenomenology
is the fact that combining the study of these puzzling
astrophysics data with the study of 
data on the time/energy structure 
of gamma-ray bursts it is possible to distinguish between
alternative Planck-length-phenomenology models.

Let me start with a brief discussion of the
experimental paradoxes analyzed in the mentioned
studies by Protheroe and Meyer\cite{aus} 
and by Kifune.\cite{kifu}
The better known paradox is the one which Kifune
(and, later, several other authors; see, {\it e.g.},
Refs.~[6,7,8])
considered from the Planck-length-phenomenology viewpoint.
This concerns the fact that
Ultra High Energy Cosmic Rays (UHECRs) 
should interact with the Cosmic Microwave Background
Radiation (CMBR) and produce pions. 
These interactions should forbid\cite{GZK} the arrival
on Earth of UHECRs 
with $E > 5 {\times} 10^{19}$eV (the GZK limit),
but several cosmic rays have been observed\cite{AgaWat} 
with nominal energies at or above $10^{20} {\pm} 30\%$ eV.
In Refs.~[5-8]
it was observed that this ``threshold anomaly"\cite{gactp2}
could be explained by assuming that the correct value
of the threshold energies is not the one that follows from 
conventional Lorentz kinematics: values of the threshold
that are consistent with the data can be
obtained using deformed rules of kinematics based 
on the Planck-length-phenomenology proposal 
of Ref.~[4].

The paradox which Protheroe and Meyer\cite{aus} (and, later, 
Refs.~[7,11])
considered from the Planck-length-phenomenology viewpoint
is somewhat analogous to the cosmic-ray paradox.
HEGRA has detected\cite{Aharonian99} high-energy photons with a 
spectrum ranging up to 24 TeV from Mk501 (Markarian 501), a BL Lac
object at a redshift of 0.034 ($\sim 157$ Mpc). 
Also these observations are puzzling since a high energy photon
propagating in the intergalactic space can interact with an IR
background photon and produce an electron-positron pair if the
product of the energies of the two photons exceeds $m_e^2$. 
Pair production should forbid
the arrival on Earth of Mk501 photons with energies
above 10 TeV, in clear conflict with the mentioned
observations of Mk501 photons with energies up to 24 TeV. 
(Using current
IR background estimates Coppi and Aharonian\cite{Coppi99} find an
optical depth of 5 for 20TeV photons from Mk501.)
In Refs.~[3,7,11]
it was observed that this Mk501 threshold anomaly
could be explained by assuming that the correct value
of the threshold energies is obtained using deformed rules 
of kinematics based on the Planck-length-phenomenology proposal
of Ref.~[4].

The UHECR and the Mk501 threshold paradoxes
clearly have common features, although they involve different
processes and different energy scales. 
There are various proposals that would provide
new-physics interpretations of either the 
UHECR paradox\cite{Olinto,Gonzalez,colgla,bertli} 
or the Mk501 paradox\cite{othermk},
but the solution based on
the Planck-length-phenomenology proposal 
of Ref.~[4]
receives some encouragement from the fact that,
as Piran and I showed recently\cite{gactp2},
there is a range of values of the deformation scale 
(which is in principle
a free parameter in the proposal 
of Ref.~[4])
that solves simultaneously both paradoxes and this range of values
is centered around the Planck length (justifying the
interpretation as a quantum-gravity effect).

While it is still quite plausible that
the correct solutions of the paradoxes might not even
involve a deformation of the thresholds
(for example, one of the proposed solutions of the UHECR paradox
evaluates threshold energies in the conventional manner
but advocates processes of decay of topological defects),
in the remainder of this lecture I take as working assumption
that the paradoxes do require a departure from the conventional 
way to evaluate threshold energies. Since I plan to discuss
a few alternative scenarios for threshold shifts it is useful
to quickly review the conventional evaluation of the threshold
in the case of head-on 
collision between a soft photon of energy $\epsilon$ and momentum $q$ 
and a high-energy particle of energy $E_1$ and momentum $\vec{p}_1$ 
leading to the production of two particles with energies $E_2$,$E_3$ and 
momenta $\vec{p}_2$,$\vec{p}_3$.  At threshold (no energy available
for transverse momenta), energy conservation 
and momentum conservation imply 
\begin{equation} 
E_1+\epsilon=E_2+E_3 
~, 
\label{econsv} 
\end{equation} 
\begin{equation} 
p_1-q=p_2+p_3~; 
\label{pconsv} 
\end{equation} 
moreover, using the ordinary 
Lorentz-invariant relation between energy and momentum
(the {\it in vacuo} dispersion relation)
\begin{equation} 
E^2=p^2+m^2 \, \, , 
\label{disprel} 
\end{equation} 
one also has the relations 
\begin{equation} 
\epsilon = q~,~~~E_i = \sqrt{p_i^2+m_i^2} 
\simeq p_i + {m_i^2 \over 2 p_i} 
~, 
\label{lirel} 
\end{equation} 
where $m_i$ denotes the mass of the particle with momentum $p_i$ and 
the fact that $p_1$ (and, as a consequence, $p_2$ and $p_3$) is a 
large momentum has been used to approximate the square root. 

The threshold value $p_{1,th}$ of the momentum $p_1$ can be identified 
with the 
requirement that the solutions for $p_2$ and $p_3$ as a function of 
$p_1$ (with a given value of $\epsilon$) that follow from 
Eqs.~(\ref{econsv}), (\ref{pconsv}) and (\ref{lirel}) should be 
imaginary for $p_1 < p_{1,th}$ and should be real  
for $p_1 \ge p_{1,th}$.   
This straightforwardly leads to the threshold equation 
\begin{equation} 
p_{1,th} \simeq {(m_2 + m_3)^2 - m_1^2 \over 4 \epsilon} 
~. 
\label{lithresh} 
\end{equation} 
 
As mentioned, the data on UHECRs and on Mk501 photons
might suggest that this standard result 
underestimates $p_{1,th}$, at least for the 
processes $p + \gamma \rightarrow p + \pi$ 
and $\gamma + \gamma \rightarrow e^+ + e^-$.
A result for the threshold different from (\ref{lithresh})
can of course be obtained by modifying the conditions
for energy-momentum conservation (\ref{econsv})-(\ref{pconsv})
and/or the dispersion relation (\ref{disprel}).
Within Special Relativity it is of course only possible
to introduce such modifications 
by assuming that propagation and scattering occur in 
presence of a background.
It has long been conjectured that the Planck length 
might characterize one such background: the ``quantum-gravity foam" 
or ``quantum-gravity medium".
The presence of the background would allow to single out a preferred
class of inertial frames, and it is therefore possible to assume,
as commonly done\cite{grbgac,gampul}, that the dispersion relation
is affected by a Planck-length deformation
while energy-momentum conservation is unmodified.

I recently showed\cite{dsr1,dsr2} 
that another phenomenologically and conceptually
consistent possibility is the one 
of a small Planck-length deformation of the
Relativity postulates
(basically fixing the Planck length as 
observer-independent ``minimum length"\cite{dsr2}).
This second scenario is of course more constrained:
the fact that there is no background field associated 
with the Planck length
(which implies that there is no preferred class of inertial
observers for the description of the Planck-length structure
of space-time) 
requires that the modifications of the 
dispersion relation
and of energy-momentum conservation are correlated
(both modifications must reflect the same transformation rules
between different inertial frames).

The Planck-length phenomenology scenario which has been
most carefully analyzed\cite{kifu,sato,gactp2}
with respect to the threshold anomalies
is the one in which the Planck length is introduced together
with an (unspecified) associated background, 
energy-momentum conservation is unmodified,
and in the high-energy ($E\simeq p$) regime
the dispersion relation is 
(in leading order in $L_p$) deformed according to
\begin{equation}
E^2 - \vec{p}^2 - m^2 \simeq  \eta E^2 \left({L_p E}\right)^\alpha
\simeq  \eta \vec{p}^2 \left({L_p E }\right)^\alpha
~,
\label{dispone}
\end{equation}
where $E$ and $\vec{p}$ are 
the energy and the (3-component) momentum of the particle, $L_{p}$ 
is the Planck length ($L_{p} \sim 10^{-35}m$), and $\alpha$ 
and $\eta$ are free parameters characterizing the deviation 
from ordinary Lorentz invariance.
Experimental tests of the predictions of (\ref{dispone})
were first proposed in
Ref.~[4], 
for the case $\alpha=1$, and 
in Ref.~[1], 
for generic $\alpha$.

Kifune\cite{kifu} considered the case $\alpha=1$
and observed that for $\eta \simeq -1$
the new threshold was consistent with data on UHECRs.
Aloisio {\it et al}\cite{ita}
observed that the determination of the UHECR threshold
and of the
gamma thresholds of distant BL Lac objects
could be used to test both the case $\alpha=1$
and the case $\alpha=2$.
Protheroe and Meyer\cite{aus}
considered the case $\alpha=1$
and observed that for $\eta \simeq -1$
the new threshold was consistent with data on Mk501 photons.
Piran and I\cite{gactp2} took as working assumption that both the 
UHECR and Mk501 paradoxes are indeed due to a threshold anomaly
and tested the consistency of this assumption 
by studying the corresponding constraints on 
the $\alpha,\eta$ parameter space.
We combined the (lower) bounds required by the working assumption
for the threshold anomalies
with the (upper) bounds on the same
$\alpha,\eta$ parameter space which are imposed by
the negative results of tests of (\ref{dispone}) based on time-of-flight 
analyses\cite{polonpap,grbgac,schaef,billetal} of photons emitted
by gamma-ray bursters and BL Lac objects.  According to (\ref{dispone}) 
one would predict
energy-dependent relative delays between the times of arrival of 
simultaneously emitted photons; in fact,
from (\ref{dispone}) it follows that the speed of photons
is energy-dependent: 
$v_\gamma = 1 + (1+ \alpha) \eta L_p^\alpha E^\alpha/2$.
We found that in order to solve both paradoxes and satisfy 
the time-of-flight upper bound
it is basically necessary to have $\alpha \simeq 1$
and $\eta \simeq -1$. This strict constraint is mostly due to
the Mk501 threshold anomaly, while the UHECR threshold anomaly
is softer and would be consistent with all values of $\alpha$
in the range $1 \le \alpha \le 2$.

The fact that the UHECR threshold anomaly
is softer than the Mk501 threshold anomaly is easily understood
by looking at the
formulas for the thresholds.
Assuming (\ref{dispone}) 
and assuming that 
energy-momentum conservation is unmodified
one finds 
\begin{equation} 
p_{1,th} \simeq {m_e^2 \over \epsilon} 
+ \eta {p_{1,th}^{2+\alpha} L_p^\alpha \over 4 \epsilon} \left(  
{1 \over 2^{\alpha}} -1 \right) 
~. 
\label{ggthresh} 
\end{equation} 
for  $\gamma + \gamma \rightarrow e^+ + e^-$
(relevant for the Mk501 paradox), and
\begin{equation} 
p_{1,th} \simeq {(m_p + m_\pi)^2 - m_p^2 \over 4 \epsilon} 
+ \eta {p_{1,th}^{2+\alpha} L_p^\alpha \over 4 \epsilon} \left(  
{m_p^{1+\alpha} + m_\pi^{1+\alpha} \over (m_p + m_\pi)^{1+\alpha}} -1 \right) 
~. 
\label{pgthresh} 
\end{equation} 
for $p + \gamma \rightarrow p + \pi$ 
(relevant for the UHECR paradox). The pair-production threshold
has simpler form because of the symmetry of the process,
but in both cases the correction depends on the momentum scale
of the process 
through $\eta p_{1,th}^{2+\alpha} L_p^\alpha/(4 \epsilon)$.
The coefficient of $\eta p_{1,th}^{2+\alpha} L_p^\alpha/(4 \epsilon)$
is somewhat different in the two cases (it is smaller in the photopion
production case) but the dominant difference comes from the fact 
that there are more than 5 orders of magnitude difference 
in $p_{1,th}$ ($p_{1,th}$ for UHECRs is more than 5 orders of magnitude
greater than $p_{1,th}$ for Mk501), and therefore
in comparing the magnitude of the corrections the dominant
factor is the magnitude 
of $\eta p_{1,th}^{2+\alpha} L_p^\alpha/(4 \epsilon)$.
Since the (tentative) evidence of threshold anomalies 
suggests threshold shifts by factors of the same order
(the Mk501 case requires a factor-$2$ 
upward shift, $10 TeV \rightarrow 20 TeV$,
while the UHECR case requires a factor-$6$ 
upward shift, $5 {\cdot} 10^{19} eV \rightarrow 3 {\cdot} 10^{20}eV$)
this leads to the conclusion that, for $|\eta| \simeq 1$, 
$\alpha$ cannot be much greater than 1 in order to explain 
the Mk501 paradox, while $\alpha$ as large as 2 could still
explain the UHECR paradox.

The consistency of the overall picture relies also on
the fact that we showed\cite{gactp2} that the region of the 
$\alpha,\eta$ parameter space that provides an explanation for
both the UHECR and the Mk501 paradoxes is also consistent with 
the mentioned negative results of 
searches\cite{polonpap,grbgac,schaef,billetal} of 
time-of-flight anomalies,
but the margin of consistency is relatively slim.\cite{gactp2}
Sentivity to 
time-of-flight anomalies
is going to improve sharply in a few years
with planned experiments
such as the GLAST gamma-ray space telescope\cite{glast},
and if these new searches still give negative results
the $\alpha,\eta$ solution of the threshold paradoxes 
(at least the one for the Mk501 paradox)
will be ruled out.

It is a significant development for Planck-length phenomenology
that these UHECR and Mk501 paradoxes admit interpretation in
terms of the scenario
in which the Planck length is introduced together
with an associated background, 
energy-momentum conservation is unmodified,
and one uses the two-parameter Planck-length deformation
of the dispersion relation (\ref{dispone}).
The experimental evidence (and theory work attempting to
interpret it) is much too preliminary to 
justify any serious hopes that we might have stumbled upon the first
manifestation of a ``quantum gravity", 
but the fact that for
the first time phenomenological models involving quantum-gravity
ideas are competing on level ground with other new-physics proposals
clearly marks the beginning of a long-waited new stage of
quantum-gravity research.
Moreover, even if the paradoxes are eventually understood in a way that
does not involve anomalous thresholds, 
the result would still be significant
since we would be able to rule out all pictures
of the short-distance (Planckian) structure of space-time
that predict anomalous thresholds. We finally do start having
a few experiments that probe the short-distance structure of space-time
with Planck-length sensitivity!

It is important to realize that the scenario
in which the Planck length is introduced together
with an associated background, 
energy-momentum conservation is unmodified,
and one uses the two-parameter Planck-length deformation
of the dispersion relation (\ref{dispone})
is not the only way in which quantum properties of
space-time may affect the evaluation of the thresholds.
Whenever one considers the possibility of 
quantum space-time properties that introduce a new length scale
(possibly the Planck length or the string length)
together with an associated background it is natural to find that
the dispersion relation is modified (reflecting the 
fact that the background allows to single out a preferred class
of inertial frames). Besides the $\alpha,\eta$ example
discussed in detail above, another scenario of this type
which has attracted considerable attention recently
is a string-theory scenario\cite{stringnc1,suss1}
in which indeed new length scales
(possibly, but not necessarily\cite{stringnc1,suss1}, 
identified with the Planck or the string length)
are introduced together with an accompanying background.
A good effective-theory description of this string-theory
scenario is obtained by describing all the new-physics
effects through space-time noncommutativity of the 
type $[X_\mu,X_\nu]= i \Theta_{\mu,\nu}$. 
The dimensionful (length-squared) parameters $\Theta_{\mu,\nu}$
are the only way in which the new background affects particle-physics
processes. 
Preliminary results\cite{suss1} appear to indicate that some particles
acquire $\Theta$-dependent corrections to the dispersion
relation $E^2 = p^2 + m^2$ that go\footnote{Of course,
the IR singularity of $(p^\mu \Theta_{\mu,\nu} p^\nu)^{-1}$
reflects the fact that the effective theory breaks down
in the IR limit.
Still, it appears\cite{suss1} that the effective theory
might be reliable for momenta that are sufficiently soft
for $(p^\mu \Theta_{\mu,\nu} p^\nu)^{-1}$ to have significant
implications.}
like $(p^\mu \Theta_{\mu,\nu} p^\nu)^{-1}$.
This $\Theta_{\mu,\nu}$ phenomenology
would of course affect the determination of the thresholds
in a way that is somewhat analogous\footnote{Besides the different
energy dependence of the effect, the $\Theta_{\mu,\nu}$ scenario
differs from the $\alpha,\eta$ scenario also because it predicts
a polarization dependence for the dispersion-relation deformation
that applies to photons. A similar polarization dependence is also
predicted by preliminary studies~\cite{gampul} of deformed dispersion
relation in Loop Quantum Gravity. However, the effect described
in Ref.~\cite{gampul} grows with energy (as in the $\alpha,\eta$ scenario)
and can therefore be distinguished from the $\Theta_{\mu,\nu}$ scenario.} 
to the $\alpha,\eta$
phenomenology discussed above. A detailed analysis of threshold anomalies
in the $\Theta_{\mu,\nu}$ scenario is now in preparation.\cite{gacsnprep}
Even before a detailed analysis one easily concludes that,
since the $\alpha,\eta$ scenario
predicts effects that increase with the particle momentum, while
the $\Theta_{\mu,\nu}$ scenario predicts effects that decrease with 
the particle momentum,  
one of the two scenarions should emerge as experimentally favoured
(or both will be ruled out), when eventually 
the UHECR and the Mk501 thresholds
will be determined/understood and experiments on time-of-flight anomalies
will be more accurate. 

As I observed in recent work\cite{dsr1,dsr2},
another logically consistent and phenomenologically viable
possibility for the Planck length to characterize space-time structure
it the one in which $L_p$ has the role of observer-independent length scale.
In the examples discussed above, the space-time-foam $\alpha,\eta$ scenario
and the stringy $\Theta_{\mu,\nu}$ scenario, space-time length scales
are introduced in a way that would allow to single out a preferred
class of inertial observers, but it is also possible\cite{dsr1,dsr2}
to attribute to $L_p$ a role in Relativity which is completely
analogous to the role of the velocity scale $c$: both $c$ and $L_p$
could be observer-independent and their role in space-time structure
would not allow to distinguish between different inertial frames.
In order to give an intuitive characterization of the implications of
(and, particularly, the constraints imposed by) 
this drastic, but compellingly simple, assumption, let me analyze
the dispersion relation (\ref{dispone}), 
in the simple case of photons ($m=0$) and $\eta=\alpha=1$, 
$E^2 - \vec{p}^2 \simeq  \vec{p}^2 {L_p E }$.
If $E$ and $\vec{p}$ transform from one to another inertial frame
according to ordinary Special Relativity, than this dispersion relation
can only be valid in one preferred class of inertial frames,
other inertial observers would not agree on the
correction term (they would attribute to $L_p$ a value that is different
from the one obtained following Planck's prescription).
It is however possible\cite{dsr1,dsr2} 
to add to the Relativity postulates the
requirement that the dispersion 
relation $E^2 - \vec{p}^2 \simeq  \vec{p}^2 {L_p E }$
is valid in all inertial frames, for fixed observer-independent value 
of $L_p$. The corresponding rules\cite{dsr1,dsr2} of transformation
of $E$ and $\vec{p}$ are of course not exactly the ones of ordinary
Special Relativity, but as long as the relative velocity between
the inertial observers is not large the deformation is very mild
(and in particular the new transformation rules reproduce
the old ones in the small relative velocity limit).

If photons satisfy the dispersion 
relation $E^2 - \vec{p}^2 \simeq  \vec{p}^2 {L_p E }$
in {\it all} inertial frames (with fixed observer-independent $L_p$),
massive particles should accordingly 
satisfy in all inertial frames a dispersion relation of the 
type $E^2 - p^2 - m^2 \simeq F(E,p;m;L_p)$,
with $F$ some function such that 
(in leading order in $L_p$) $F(E,p;0;L_p) = \vec{p}^2 {L_p E }$.
For the simplest possibility, $m$-independent $F$
({\it i.e.} $F(E,p;m;L_p) = \vec{p}^2 {L_p E }$ for all $m$),
I already verified the consistency of the
overall scenario.\cite{dsr1} 
Work is now in progress\cite{dsr3inprep}
for other forms of $F$.
The case of $m$-independent $F$ is however sufficient to illustrate
the implications of assuming that the deformed dispersion
relation is valid in all inertial frames (rather than
only in a specific class of inertial frames).
In fact, the same 
dispersion relation $E^2 - p^2 - m^2 \simeq \vec{p}^2 {L_p E }$
which causes large threshold anomalies when considered in the
sense 
of Ref.~[4] 
(valid only in certain inertial frames),
generates only very small threshold anomalies when it is assumed
that it be valid in all inertial frames.
The point is that if $E^2 - p^2 - m^2 \simeq \vec{p}^2 {L_p E }$
is valid in all inertial frames, then, as mentioned, the
transformation rules between different inertial frames must
be modified and the conditions for
energy-momentum conservation must also be accordingly modified.
The new conservation rules\cite{dsr1} are not very different
from ordinary energy-momentum conservation (just like the 
dispersion relation has a 
Planck-length-suppressed deformation
there is of course a corresponding
Planck-length-suppressed deformation of the conservation rules)
but they combine with the deformed dispersion relation
in such a way that there is a nearly exact cancellation of
deformation terms for the threshold, so that the threshold 
anomaly is much smaller than in the case in which 
ordinary energy-momentum conservation is assumed.
Therefore experiments indicating explicitly a relatively large threshold
anomaly would disfavour this scheme (but large threshold anomalies
are not the only way to interpret the UHECR and Mk501 data: the observations
could also be explained with sizeable anomalies just above threshold,
where the mentioned cancellation is not efficient). 

The implications of the deformed dispersion 
relation $E^2 - p^2 - m^2 \simeq \vec{p}^2 {L_p E }$
depend  strongly on whether it is assumed to hold
in all inertial frames (or just in one class of them)
also for what concerns the mentioned time-of-flight anomalies.
Searches\cite{glast} of time-of-flight
anomalies should expect a dependence on the velocity of the
source in the case in which the deformed dispersion relation
singles out a preferred class of inertial frames (the velocity
of the emitting galaxy with respect to the preferred frame
would have absolute physical meaning in that scenario),
while such a dependence should be absent if the Relativity
postulates are consistently modified in such a way\cite{dsr1}
that the deformed dispersion relation holds in all inertial frames.

Combining astrophysics data on possible threshold anomalies 
and the results of searches of time-of-flight anomalies
we can therefore not only rule out several candidate short-distance
(quantum) structures of space-time, but (if any anomalies are
actually found/confirmed) we should be able to distinguish between
different scenarios in which the Planck length
is introduced together with an accompanying background
(and a preferred class of inertial frames),
{\it e.g.} we should be able to
distinguish between the $\alpha,\eta$ scenario
and the $\Theta_{\mu,\nu}$ scenario with preferred class
of inertial frames, and we should also be able to distinguish
between the possibility that 
the Planck scale is introduced together with an accompanying
preferred class of inertial frames and
the possibility that 
the Planck length is introduced as an observer-independent
characteristic of space-time structure,
{\it e.g.} we should be able to 
distinguish between the case in which the $\alpha,\eta$ dispersion
relation is valid only in a preferred class of inertial frames
and the case in which the $\alpha,\eta$ dispersion
relation is valid in all inertial frames.

\section*{Acknowledgments}
These notes are based both on recent independent 
work\cite{dsr1,dsr2}
and on recent work in collaboration with Tsvi Piran\cite{gactp2}
and with Luisa Doplicher and Soonkeon Nam\cite{gacsnprep}.
Conversations with Luisa,
Soonkeon and Tsvi, and with Roberto Aloisio, Aurelio Grillo 
and Jurek Kowalski-Glikman, are gratefully acknowledged.

\end{document}